\documentclass[11pt]{article}

\usepackage[english,francais]{babel}

\usepackage[latin1]{inputenc}
\usepackage{epsfig,graphicx}
\usepackage{latexsym}

\newfont{\ensmathquatorze}{msbm10 scaled 1400}
\newfont{\ensmathonze}{msbm10 scaled 1100}
\newfont{\ensmathdix}{msbm10}
\newfont{\ensmathneuf}{msbm10 scaled 833}
\newfont{\ensmathhuit}{msbm10 scaled 694}
\newfam\ensmathfam                        
\textfont\ensmathfam=\ensmathonze        
\scriptfont\ensmathfam=\ensmathdix       
\scriptscriptfont\ensmathfam=\ensmathhuit

\def\be{\begin{equation}}
\def\ee{\end{equation}}
\def\bea{\begin{eqnarray}}
\def\eea{\end{eqnarray}}

\def\d{{\rm d}}
\def\sqg{\sqrt{2  \gamma}}
\def\sqx{\sqrt{2  ( \gamma + \xi )}}
\def\ga{\gamma }
\def\gx{\gamma + \xi }
\def\go{ {\cal G}_0 }
\def\gd{ {\cal G}_D }
\def\fp{\displaystyle }

\begin{document}

\selectlanguage{english}

\title{ The Last Passage Problem on Graphs }

\author{ Jean Desbois$^1$    and Olivier B\'enichou$^2$   }

\maketitle	

{\small
\noindent
$^1$ Laboratoire de Physique Th\'eorique et Mod\`eles Statistiques.
Universit\'e Paris-Sud, B\^at. 100, F-91405 Orsay Cedex, France.

\noindent
$^2$ Laboratoire de Physique de la Mati\`ere Condens\'ee, Coll\`ege de France,
 11, place M. Berthelot, 75231 Paris Cedex 05, France.
}

\begin{abstract}

We  consider a brownian motion on a general graph, that starts at time $t=0$
 from some vertex $O$ and stops at time $t$ somewhere on the graph.
 Denoting by $g$ the last time when $O$ is reached, we establish a simple
 expression for the Laplace Transform, $L$, of the probability density
 of $g$. We discuss this result for some special graphs like star,
  ring, tree or square lattice.
  Finally, we show that $L$ can also be expressed in terms of primitive orbits
   when, for any vertex, all the exit probabilities are equal.

\end{abstract}

\vskip2cm

\noindent
Around the years 1930, P Levy \cite{plevy}  got several
 arc-sine laws concerning the  1D Brownian Motion on  an infinite line.
 Let us consider such a  process starting at $t=0$ from the origin $O$
 and stopping
  at time $t$ and   denote by $g$ the last time
  when $O$ is reached.
  The second  arc-sine law discovered by
 Levy concerns the probability law of $g$. It can be stated as follows:  

\bea
           P(g<u) & = & \frac{2}{\pi } \; \arcsin 
	    \sqrt{ \frac{u}{t}   }                     \label{levy1} \\
 \mbox{with the probability density } \qquad 
       {\cal P}_t(u) \equiv \frac{\d  P(g<u) }{ \d u}
       & = & \frac{1}{\pi } \; \frac{1}{ \sqrt{u (t-u) }    }
    \label{levy2}
\eea

\vskip.5cm
\noindent
In particular, the Laplace transform of ${\cal P}_t(u)$ is written:

\be\label{levy3}
 \int_0^{\infty } \d t \; e^{-\gamma t} \int_0^t \d u \; {\cal P}_t(u)
 \; e^{-\xi u}  \; = \; \frac{1}{\sqrt{\gamma (\gamma +\xi)}}
\ee

\vskip.5cm

\noindent
Levy also proved that the same law occurs when $g$ represents the time spent
 in the region $(x>0)$ (first arc-sine law).

\vskip.2cm

\noindent
The infinite line with only one point, $O$, specified,
  can be viewed as a kind of very simple graph consisting in one
   vertex, $O$, and two semi-infinite lines originating from $O$. 
  
\vskip.3cm

Our goal in this Letter is to get an analogue of the second arc-sine law
 (\ref{levy3})  but  for a  quite general graph.

\vskip.3cm

\noindent
Let us recall that graphs properties have interested for many years
 physicists  \cite{mont} 
 as well as mathematicians   \cite{roth}.
 For instance, the knowledge of  spectral properties
 of the Laplacian operator is useful to understand superconducting networks
  \cite{superc} as well as  organic molecules \cite{organ}.
  We also know that spectral determinants play a central role in the
  computation of weak localization corrections \cite{mont1}.
   Recently, graphs have also been used in the context of nonequilibrium
   statistical physics (see  \cite{dhar} and references therein) and
  also in the field of quantum chaos \cite{ko}.
  
\vskip.2cm  

\noindent
The study  of Brownian motion on graphs is also an active area of
research. For instance, occupation times \cite{myor, jd1} or local time
 \cite{cdm} distributions have been recently computed for general graphs.

\vskip.6cm  

\noindent
So, let us begin by setting the notations that will be used throughout this
 Letter. We  consider  a  
 graph  made of $V$ vertices,
   numbered from $0$ to $V-1$, linked by $B$ bonds of finite lengths.
   The coordination  number 
of vertex $\alpha$ is   $m_{\alpha }$ 
 ($\sum_{\alpha =0 }^{V-1} m_{\alpha }= 2 B$).

\vskip.2cm

\noindent
On each bond [$\alpha\beta$], of length
     $l_{\alpha\beta }$, we define the coordinate $x_{\alpha\beta }$ that
     runs from $0$ (vertex $\alpha$) to $l_{\alpha\beta }$ (vertex $\beta$).
 Sometimes we will also use the coordinate 
 $x_{\beta\alpha } \equiv l_{\alpha\beta } - x_{\alpha\beta } $.
 The set of coordinates  $\{ x_{\alpha\beta } \} $ is simply noted $x$.  

\vskip.2cm

\noindent
An arc  ($\alpha\beta$) is defined as the oriented bond
  from $\alpha$ to $\beta$.
 Each bond  [$\alpha\beta$]
 is therefore associated  with two arcs
    ($\alpha\beta$) and  ($\beta\alpha$). In the sequel, we will consider the
  following ordering of the $2B$ arcs: 
 $ (0\mu_1)(0\mu_2)\ldots (0\mu_{m_0})\ldots (\alpha\beta_1)\ldots
  (\alpha\beta_i)\ldots  (\alpha\beta_{m_\alpha })\ldots  $

\vskip.6cm

\noindent
Now, let us consider   a brownian motion starting at  some
 vertex $O$ (label $0$) of the graph at time $t=0$ and stopped at
  time $t$. Let $g$ be the last time when $O$ is visited.
  The probability, $P_t(g<u)$ can be written:
  
\be\label{j100}  
P_t(g<u) = \int_{{\rm Graph}} \d x' \;  \d x \; 
 G_0(0,0;x,u) \; G_D(x,u;x',t)
\ee
where $G_0$ is the free propagator on the graph and $G_D$ is the propagator
 with Dirichlet boundary conditions in $O$. (\ref{j100}) simply
  expresses the fact that the  brownian particle will not hit $O$
 between the times $u$ and $t$. Let us define the probability
  density $\fp  {\cal P}_t(u) \equiv \frac{\d  P(g<u) }{ \d u}$. 
 In the sequel, we will be especially interested in
  the Laplace transforms:
 
\bea
{\cal L} & \equiv &           
\int_0^{\infty } \d t \; e^{-\gamma t} \int_0^t \d u \; P_t(g<u)
 \; e^{-\xi u}     \label{LT1}     \\
  L & \equiv &     
 \int_0^{\infty } \d t \; e^{-\gamma t} \int_0^t \d u \; {\cal P}_t(u)
 \; e^{-\xi u} \;  =  \;
  \frac{1}{\gx } + \xi {\cal L}
  \label{LT2}     
\eea

\vskip.2cm

\noindent
With (\ref{j100}), we see that:
\be\label{j101}
{\cal L}= \int_{{\rm Graph}} \d x' \d x 
 \left\langle   0   \left\vert \frac{1}{H+ \gx }
  \right\vert  x \right\rangle
\left\langle   x   \left\vert \frac{1}{H_D+ \ga }
  \right\vert  x' \right\rangle \equiv 
 \int_{{\rm Graph}} \d x' \d x \; \go (0,x;\gx ) \; \gd (x,x';\ga )
\ee
On the bond  $[\alpha\beta ]$, $H$ and $H_D$ are equal to 
  $ \fp - \frac{1}{2}  \frac{\d ^2}{\d x_{\alpha\beta } ^2 } $ and we define:
\bea
\go ^{(\alpha\beta )}= \lim_{x_{\alpha\beta }\to 0 }
  \go (0, x_{\alpha\beta }; \gx) & \qquad ; \qquad & 
 \go ^{' \; (\alpha\beta )} =\lim_{x_{\alpha\beta }\to 0 }
\frac{  \d \go (0, x_{\alpha\beta }; \gx) }
{\d  x_{\alpha\beta } }   \label{defin1} \\
\gd ^{(\alpha\beta )}(x)= \lim_{x'_{\alpha\beta }\to 0 }
  \gd (x, x'_{\alpha\beta }; \ga) & \qquad ; \qquad & 
 \gd ^{' \; (\alpha\beta )}(x) =\lim_{x'_{\alpha\beta }\to 0 }
\frac{  \d \gd (x, x'_{\alpha\beta }; \ga) }
{\d  x'_{\alpha\beta } }   \label{defin2} 
\eea

\vskip.3cm

\noindent
The integration over $x'$ in (\ref{j101}) is easily performed with the help 
 of the equation 
\be\label{j102}
 \left( -\frac{1}{2} \; \frac{\d ^2}{\d x^{'2}_{\alpha\beta }  } \; + \; 
 \gamma 
 \right) \gd (x,x'_{\alpha\beta };\ga ) =\delta(x - x'_{\alpha\beta })
\ee
We get 
\be\label{j103}
\int_{{\rm Graph}} \d x' \; \gd (x,x'; \ga )
 = \frac{1}{\ga } - \frac{1}{2 \ga } \; \sum_{i=1}^{m_0}
   \gd ^{' \; (0 \mu_i )}(x) 
\ee

\vskip.3cm

\noindent
To go further, we have to specify  the behaviours of the resolvants
 $\go$ and $\gd$ in the neighbourhood of all the vertices. 
   
\vskip.1cm

\noindent
Let us consider
 some vertex $\alpha$ with its nearest neighbours $\beta_i$,
 $i=1,2,\ldots  m_{\alpha }$, on the graph. Suppose that the brownian
 particle reaches $\alpha$. It will come out towards $\beta_i$ with some
 exit  probability $p_{\alpha\beta_i }$. This implies
 the following   equations  to be satisfied by $\go $:
 
\be\label{resolv}
\frac{\go ^{(\alpha\beta_1 )}}{p_{\alpha\beta_1}} =
\frac{\go ^{(\alpha\beta_2 )}}{p_{\alpha\beta_2}} =
\ldots =
\frac{\go ^{(\alpha\beta_{m_\alpha } )}}{p_{\alpha\beta_{m_\alpha }}}
 \equiv f_{\alpha } \quad , \quad \forall \alpha
\ee

\vskip.2cm

\noindent
Remark that the resolvant will be continuous in vertex $\alpha $
 only when the particle exits from $\alpha $ with the same probability in all
  the directions. 
 
\vskip.2cm

\noindent
Moreover, current conservation implies:

\bea
\sum_{i=1}^{m_{\alpha } }  \go ^{' \; (\alpha\beta_i )}
 & = & 0 \quad , \ \mbox{if} \ \alpha \ne 0 \label{j110}   \\
\sum_{i=1}^{m_0 }  \go ^{' \; (0 \mu_i )}
 & = & -2 \label{j111} 
\eea

\vskip.2cm

\noindent
On the link
  $[\alpha\beta ]$,  $\go (0,x_{\alpha\beta }; \gx)$ must satisfy: 

\be\label{j112}
 \left( -\frac{1}{2} \; \frac{\d ^2}{\d x_{\alpha\beta } ^2 } \; + \; 
 \gx 
 \right) \go (0,x_{\alpha\beta }; \gx ) =0
\ee
with the solution:

\be\label{j113}
\go (0,x_{\alpha\beta }; \gx ) =
\go ^{(\alpha\beta )} \; 
 \frac{ \sinh \sqx (l_{\alpha\beta } - 
 x_{\alpha\beta })}
     { \sinh \sqx  l_{\alpha\beta } } +
 \go ^{(\beta\alpha ) } \; 
\frac{ \sinh \sqx x_{\alpha\beta } }
     { \sinh \sqx   l_{\alpha\beta } } 
\ee

\vskip.1cm
\noindent
So, we deduce:
\be\label{j114}
 \go ^{' \; (\alpha\beta ) } =
  \sqx \left(
 -  c_{\beta\alpha } (\gx ) \; \go ^{(\alpha\beta ) } +
 s_{\alpha\beta } (\gx) \; \go ^{(\beta\alpha ) } \right)  
\ee
with:
\bea
c_{\alpha\beta } (\gx ) &=&  
 \coth \sqx l_{\alpha\beta } 
 = c_{\beta\alpha }(\gx ) \label{j115} \\
  s_{\alpha\beta } (\gx) &=&
  \frac{ 1 }{
  \sinh \sqx  l_{\alpha\beta }} =  s_{\beta\alpha }(\gx )
   \label{j116}
\eea

\noindent
Those considerations allow to write eqs.(\ref{resolv}, \ref{j110}, \ref{j111})
in a matrix form:

\be\label{mat1}
M (\gx) \; f \; = \frac{1}{ \sqx } \; l
\ee
where $f$ and $l$ are two $(V \times 1) $ vectors. The components of $f$ are
 the quantities
 $f_{\alpha }$ and  for the components of $l$ we have 
  $l_j = 2 \delta_{j 0}$.

\vskip.2cm

\noindent
$M ( \gx )$ is a  $(V \times V)$  vertex matrix with the non vanishing elements
($\alpha$ runs from $0$ to $V-1$):

\bea
   M( \gx )_{\alpha\alpha }   &= & \sum_{i=1}^{m_{\alpha }} 
   p_{\alpha\beta_i }  \;    c_{\alpha\beta_i } ( \gx )   \label{j117}   \\
   M( \gx )_{\alpha\beta }   &= & - \; p_{\beta\alpha } \;   
   s_{\alpha\beta }  ( \gx )   
   \qquad  \mbox{if } \ [\alpha\beta ] \  \mbox{is  a bond}  \label{j118} 
   \\
     &= & 0  \qquad  \mbox{ otherwise }
\eea

\vskip.2cm

\noindent
Similar considerations hold for $\gd (x,x'; \ga )$ and we get:

\bea
\frac{\gd ^{(\alpha\beta_1 )} (x) }{p_{\alpha\beta_1}} =
 \ldots =
\frac{\gd ^{(\alpha\beta_{m_\alpha } )} (x) }{p_{\alpha\beta_{m_\alpha }}}
 &\equiv & F_{\alpha } (x) \quad , \quad \mbox{if } \ 
 \alpha \ne 0 \label{j119} \\
 \gd ^{( 0 \mu_i )} (x) & = & 0 \quad ; \quad i=1,\ldots ,m_0 \label{j120} \\ 
 \sum_{i=1}^{m_{\alpha } }  \gd ^{' \; (\alpha\beta_i )} (x)
 & = & 0 \quad , \quad \mbox{if } \ \alpha \ne 0 \label{j121}   \\
\mbox{and the matrix equation } \quad M_D ( \ga ) F(x) & = &
   \frac{1}{ \ga } \; l_D (x) 
 \label{j122} 
\eea
$M_D$ is the matrix $M$ with the first line and first column deleted.
 $F(x)$ and $l_D(x)$ are $(V-1)\times 1$ vectors. The components of
  $F(x)$ are the quantities $F_\alpha (x)$, $\alpha =1,\ldots ,V-1$. The
  nonvanishing components of $l_D(x)$ are defined as follows. 

\vskip.1cm

\noindent
If $x \in [ab]$ with $a\ne 0$ and $b\ne 0$ then 
\bea
  \left( l_D(x_{ab}) \right)_a & = & s_{ab}( \ga ) \sinh ( \sqg \; x_{ba})
  \label{j123} \\
 \left( l_D(x_{ab}) \right)_b & = & s_{ab}( \ga ) \sinh ( \sqg \; x_{ab})
  \label{j124} 
\eea
If $x \in [0 \mu_j]$  then 
\be\label{j125} 
\left( l_D(x_{0 \mu_j} \right)_{\mu_j}  =  s_{0\mu_j }( \ga ) 
 \sinh ( \sqg  \; x_{0 \mu_j})
\ee

\vskip.1cm

\noindent
Performing the integration over $x$ in (\ref{j101}), we finally get the
 following simple expression for $L$:
\be\label{j126} 
 L=\frac{ 1  }{ \sqrt{ \ga ( \gx )}  } 
 \frac{ \left(   M^{-1}( \gx )   \right)_{00} }  
      { \left(   M^{-1}( \ga )   \right)_{00} } 
\ee

\vskip1cm

\noindent
This is the central result of this paper.

\vfill\eject

\noindent
Remark that, for the special case when $\go $ and $\gd $ are continuous
 Green's functions, we can recast (\ref{j126}) in the form
{\footnote {
This is readily done with the help of the Dirichlet Green's function:
$$ \fp
\gd (x,x' ; \ga) = \go (x,x' ; \ga) - 
\frac{ \go (x,0 ; \ga) \go (0,x' ; \ga)   }
     { \go (0,0 ; \ga) }
$$

\noindent
Using  completeness and also the relationship
$$ \fp
 \int_{\mbox{Graph }} \d x' \; \go (x,x' ; \lambda ) =\frac{1}{ \lambda }
$$
all the integrations in   (\ref{j101}) can be performed, leading to 
 (\ref{j126bis}).
}}

\be\label{j126bis} 
 L=\frac{ 1  }{ \ga   } \;
 \frac{  \go (0,0; \gx )    }  
      {  \go (0,0; \ga )    }  
\ee

\vskip.9cm

\noindent
To go further and deduce from (\ref{j126}) some 
 general properties of the density ${\cal P}_t(u)$,
 let us consider the following identity 
\be\label{j127}
\int_0^\infty \d t \; e^{- \ga t} \int_0^t \d u \; e^{-\xi u} r(u) s(t-u)
 = {\cal R}( \gx ){\cal S}( \ga )
\ee
where $ {\cal R} $ and $ {\cal S} $ are the Laplace Transforms of $r$ and 
 $s$. 

\vskip.1cm

\noindent
Setting 
 $\fp {\cal R}(p) = \frac {\left( M^{-1}( p ) \right)_{00}}{\sqrt{p}}$
 and 
 $\fp {\cal S}(p) = \frac {1}{\sqrt{p} \; \left( M^{-1}( p ) \right)_{00} }$
  \ in (\ref{j127}),  we realize from (\ref{j126}) and (\ref{LT2})
  that the probability density factorizes
   in the following way:
\be\label{j128}
 {\cal P}_t(u) \; = \; r(u).s(t-u)
\ee

\vskip.1cm

\noindent
This result holds for any graph.

\vskip.1cm

\noindent
It is now easy to analyze the limiting behaviours of $ {\cal P}_t(u) $ when
 $u \to 0^+$ and when $u \to t^-$.
 
\vskip.1cm

\noindent
In the first case, we can write $ {\cal P}_t(u) \approx r(u) s(t)$.
 Moreover, for large $p$ values, $M(p) \to 1$ and 
 $\fp {\cal R}(p) \sim \frac{1}{\sqrt{p}}$. So
\be\label{j129} 
{\cal P}_t(u) \sim \frac{1}{\sqrt{\pi \; u}} \; s(t)
 \quad , \quad \mbox{when } \ u \to 0^+ 
\ee
On the other hand,  when $u \to t^-$, $ {\cal P}_t(u) \approx r(t) s(t-u)$.
 For large $p$ values,
 $\fp {\cal S}(p) \sim \frac{1}{\sqrt{p}}$ and finally 
\be\label{j130} 
{\cal P}_t(u) \sim \frac{1}{\sqrt{\pi \; (t-u)}} \; r(t)
 \quad , \quad \mbox{when } \ u \to t^- 
\ee

\vskip.7cm

\noindent
Let us now develop some examples.

\vskip.7cm

\begin{figure}[!h]
\begin{center}
\includegraphics[scale=.3,angle=0]{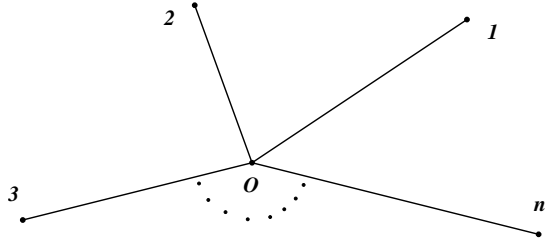}
\caption{ a star-graph with $n$ legs of lengths $l_{0i}$,
 $i=1,\ldots n    $, originating from $O$. }
\end{center}
\end{figure}

\vskip.4cm

\noindent
For the graph of Figure 1, we get:

\bea
L & = & \frac{1}{\sqrt{ \ga ( \gx )}} 
\frac{ \sum_{i=1}^n p_{0i} \; \mbox{th}(\sqg l_{0i} )  }
    { \sum_{i=1}^n p_{0i} \; \mbox{th}(\sqx l_{0i} )  } \label{j131} \\
  & \to &  \frac{1}{\sqrt{ \ga ( \gx )}} \quad, \quad
   \mbox{when } l_{0i} \to \infty \ , \ i=1,\ldots ,n \label{j132}
\eea
When all the legs become infinite, $L$ does not depend on the exit
 probabilities from $O$, $p_{0i}$, because in that case all the legs are
 equivalent. The same behaviour occurs for a finite symmetric star
  ($l_{01}=\ldots =l_{0n} \equiv l$) where we get
\be\label{j133} 
L  =  \frac{1}{\sqrt{ \ga ( \gx )}} 
\frac{ \mbox{th}(\sqg l )  }
    {  \mbox{th}(\sqx l )  } 
\ee
that leads to the density
\bea 
\fp
{\cal P}_t(u) & = & \frac{1}{ \pi \sqrt{u(t-u)}}
 \left(  \sum_{p= - \infty }^{ + \infty } (-1)^p e^{- \frac{2 p^2 l^2}{t-u}}
 \right)
\left(  \sum_{q= - \infty }^{ + \infty }  e^{- \frac{2 q^2 l^2}{u}}
 \right) \label{j134} \\
   & = & \frac{1}{ 2 l^2}
 \left(  \sum_{m= - \infty }^{ + \infty }  
  e^{- \frac{ ( 2 m - 1 ) ^2 \pi^2 (t-u) }{ 8 l^2 }}
 \right)
\left(  \sum_{n = - \infty }^{ + \infty } 
e^{- \frac{ n^2 \pi^2 u}{2 l^2}}
 \right) \label{j1344}
\eea
This last formula allows for studying the large $u$ and $t$  behaviours
  of ${\cal P}_t(u)$. For instance:
  
\be\label{j1345}  
{\cal P}_t(u) \sim \frac{1}{l^2} \; e^{-\frac{\pi^2 (t-u)}{ 8 l^2}} 
 \quad \mbox{ when } u\gg l^2 \ \mbox{ and } \ (t-u) \gg l^2
\ee

\vskip.6cm

\noindent
In Figure 2, we consider a ring where the vertices $O$ and $1$ are linked by
  two bonds of lengths $l_{01}^{(1)}$ and  $l_{01}^{(2)}$
  ($ l \equiv  l_{01}^{(1)} +  l_{01}^{(2)} $).   
  The exit probabilities from $O$ (resp. $1$) are
  $p_{01}^{(1)}$ and $p_{01}^{(2)}$
   (resp. $p_{10}^{(1)}$ and $p_{10}^{(2)}$) and we define
 \  $\fp c_{01}^{(i)} ( \ga ) = \coth \sqg l_{01}^{(i)}$,
 \  $\fp s_{01}^{(i)} ( \ga ) = \frac{1}{\sinh \sqg l_{01}^{(i)}}$,
 \  $i=1,2$. 

\vskip.5cm

\begin{figure}[!h]
\begin{center}
\includegraphics[scale=.3,angle=0]{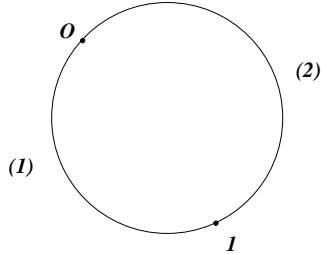}
\caption{ a ring with two  vertices, $O$ and $1$,  linked by 
 two bonds $(1)$ and $(2)$. }
\end{center}
\end{figure}

\vskip.5cm

\noindent
Adding intermediate vertices on each bond, we easily show that the
 matrix $M$ writes:

$$
 M =\left(
\begin{array}{cc} 
   \sum_{i=1}^2 p_{01}^{(i)}  c_{01}^{(i)}
    & - \sum_{i=1}^2 p_{10}^{(i)}  s_{01}^{(i)}   \\
    &           \\
    -  \sum_{i=1}^2 p_{01}^{(i)}  s_{01}^{(i)}       &
    \sum_{i=1}^2 p_{10}^{(i)}  c_{01}^{(i)}
\end{array}
\right)
$$

\vskip.3cm

\noindent
When $p_{10}^{(1)} = p_{10}^{(2)} = 1/2$, $L$ does not depend on
 $p_{01}^{(i)}  $ and we get:
 
\be\label{j133b} 
L  =  \frac{1}{\sqrt{ \ga ( \gx )}} 
\frac{ \mbox{th}(\sqg \frac{l}{2} )  }
    {  \mbox{th}(\sqx \frac{l}{2} )  } 
\ee
In that case, the ring can be viewed as a star with two legs of equal
 lengths, $l/2$, glued at their endpoints. Thus, (\ref{j133b})
 is simply (\ref{j133}) where $l$ has been replaced by $l/2$.

\vskip.3cm

\noindent
Remark, however, that we should obtain a quite different result when taking
  $p_{01}^{(1)} = p_{01}^{(2)} = 1/2$ :
\be\label{j133c} 
L  =  \frac{1}{\sqrt{ \ga ( \gx )}} 
\frac{ \mbox{coth}(\sqx \frac{l}{2} ) + F( \gx ) }
    {  \mbox{coth}(\sqg \frac{l}{2} ) + F( \ga ) } 
\ee
with 
\be\label{j133d}
\fp F( \ga ) = \frac{\left(   p_{10}^{(1)} - p_{10}^{(2)}  \right)
 \sinh \sqg \left(   l_{01}^{(2)} - l_{01}^{(1)}  \right)}
 { \cosh \sqg l \; - \; 1 }
\ee

\vskip.7cm

\noindent
Let us now turn to a brownian motion starting at the root $O$ of a regular
 tree of coordination number $z$ and depth $n$ (see Figure 3 part a) for
  $z=3$ and $n=2$). We suppose that all the links have the same length $l$
 and also that 
\bea
 p_{\alpha \beta_i }  =  \frac{1}{m_\alpha } & = & \frac{1}{z} 
 \quad \mbox{for any vertex inside the graph} \\
   & = & 1  \quad \mbox{ otherwise     }
\eea

\vskip.6cm

\begin{figure}[!h]
\begin{center}
\includegraphics[scale=.5,angle=0]{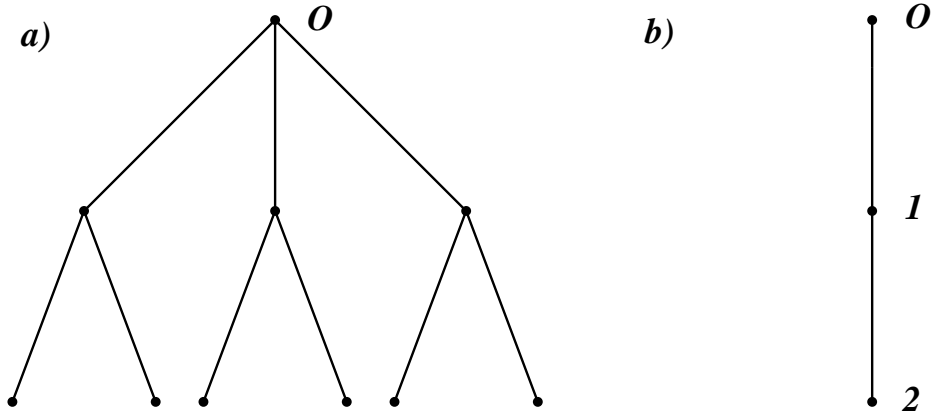}
\caption{ part a) a tree with coordination number
 $z=3$ and depth $n=2$; part b) the equivalent linear graph with exit
 probabilities given in (\ref{j137}) and (\ref{j138}).
 }
\end{center}
\end{figure}

\vskip.6cm

\noindent
For the last passage problem in $O$ this tree is equivalent to a $(n+1)$
 linear graph with the exit probabilities
\bea
p_{01} & = & 1 = p_{n,n-1} \label{j137} \\
p_{i,i-1} & = & \frac{1}{z} \quad , \quad 
 p_{i,i+1} =  1 - \frac{1}{z} \quad , \quad i=1,\ldots ,n-1 \label{j138}
\eea

\vskip.3cm

\noindent
With $\fp c( \ga ) = \coth \sqg l$ and 
  $\fp s( \ga ) = \frac{1} { \sinh \sqg l }$, we can write the relations
\bea
\mbox{det} M( \ga ) & = &  c( \ga ) \; \det M_n ( \ga )
 - \frac{1}{z} (s( \ga ))^2  \;  \det M_{n-1} ( \ga ) \label{j139} \\
\det M_{n-k}( \ga ) & = &  c( \ga ) \; \mbox{det} M_{n-k-1} ( \ga )
 + \frac{1}{z}\left(\frac{1}{z}-1   \right)
 (s( \ga ))^2 \; \det M_{n-k-2} ( \ga ) \ , \ 0\le k \le n-3 \label{j140}
 \\
\det M_2( \ga ) & = &   1 +  \frac{1}{z}(s( \ga ))^2  \label{j141}
 \\
\det M_1( \ga ) & = & c( \ga )       \label{j142}
\eea
$ M_{n-k}$ is the matrix $M$ where the $(k+1)$ first lines and first
 columns have been deleted. Recursion relations can also be written for
  $\fp \left( M^{-1} \right)_{00} \equiv \frac {\mbox{det} M_n}{\mbox{det} M}$.
   In the limit $n \to \infty$, we get:
\be\label{j143}   
L= \frac{1}{\sqrt{ \ga ( \gx )}} 
\frac{ c( \ga ) \left(1- \frac{2}{z}  \right) + \sqrt{
 1+  \left(1- \frac{2}{z}  \right)^2 (s ( \ga ))^2 }}
{ c( \gx ) \left(1- \frac{2}{z}  \right) + \sqrt{
 1+  \left(1- \frac{2}{z}  \right)^2 (s ( \gx ))^2 }}
\ee
As expected, $z=2$ leads to the second arc-sine Levy's law.

\vskip.6cm

\noindent
As a final example, let us consider an infinite 2D square lattice of stepsize
 $l$  with exit probabilities  equal
  to $1/4$ in any vertex. A standard computation based on the tight binding
  model \cite{ECONO} leads to:
\be\label{2dsql}  
L \; = \; \frac{1}{\sqrt{ \ga ( \gx )}} \;  
\frac{ \mbox{th} \sqx l }{ \mbox{th} \sqg l } \; 
\frac{  K \left( \frac{1}{ \cosh  \sqx l  }  \right)  }
{  K \left( \frac{1}{ \cosh  \sqg l  }  \right)  }
\ee
$K( \lambda )$ is a complete elliptic integral of first kind \cite{GRAD}.

\vskip1cm

\noindent
To complete this work, let us come back to a general graph 
 and choose equal exit probabilities in each vertex:
\be\label{j144}
\forall \; \alpha \quad , \quad p_{\alpha \beta_i} =
 \frac{1}{m_\alpha } \quad ; \quad i=1,\ldots , m_\alpha
\ee
In that case, $L$ can be expressed in terms of primitive orbits on the graph. 
 Recall that an orbit  $\widetilde{C}$ 
  is said to be primitive if it cannot be decomposed as
 a repetition of any smaller orbit. From \cite{jd2}, we get:  
\be\label{j145}
 \prod_{\widetilde{C}} \left( 1- \mu(\widetilde{C} )
 e^{-\sqg l(\widetilde{C} )} \right)=
 \left[ \;  \prod_{[\alpha\beta ]}\left( 
 1-  e^{- 2 \sqg l_{ \alpha \beta }}   \right) \;  \right] \; \det M( \ga )
\ee
The product on the left hand-side is taken over all the
 primitive orbits $\widetilde{C}$ of the graph. $l(\widetilde{C} )$
 is the length of $\widetilde{C}$ and the weight $ \mu(\widetilde{C} ) $
 is the product of all the reflection -- or transmission --
  factors along $\widetilde{C}$.
 (An orbit is a closed succession of arcs \
 $\ldots  (\tau\alpha )(\alpha\beta  )  \ldots $ and
  a reflection (transmission) occurs in $\alpha $
  if $\tau =\beta $   (  $\tau \ne \beta $). The corresponding factors are
   $\fp \left(  \frac{2}{m_\alpha } - 1  \right)$ for a reflection 
    and   $\fp \left(  \frac{2}{m_\alpha }  \right)$ for a transmission).
  
\vskip.1cm

\noindent
Now, imposing Dirichlet boundary conditions in $O$, we  get, still with
 \cite{jd2}:  
\be\label{j146}
 \prod_{\widetilde{C}} \left( 1- \mu^{(0)}(\widetilde{C} )
 e^{-\sqg l(\widetilde{C} )} \right)=
 \left[ \;  \prod_{[\alpha\beta ]}\left( 
 1-  e^{- 2 \sqg l_{ \alpha \beta }}   \right) \; \right]
  \; \det M_D( \ga )
\ee
$\mu^{(0)}(\widetilde{C} )$ is the same as $\mu(\widetilde{C})$
 except for orbits visiting $O$ where the reflection and transmission
  factors are respectively $-1$ and $0$.

\vskip.1cm

\noindent
It is now easy to realize that the ratio
 $\fp \frac{\det M_D( \ga )}{ \det M( \ga )}$ only involves the primitive
 orbits $ \widetilde{C_0} $ that visit $O$ at least once. Thus:
\be\label{j150} 
\left(  M^{-1} ( \ga ) \right)_{00} = 
\frac{  \prod_{\widetilde{C_0}} \left( 1- \mu^{(0)}(\widetilde{C_0} )
 e^{-\sqg l(\widetilde{C_0} )} \right)  }
{  \prod_{\widetilde{C_0}} \left( 1- \mu(\widetilde{C_0} )
 e^{-\sqg l(\widetilde{C_0} )} \right)  }
\ee
Finally:
\be\label{j151} 
L=\frac{1}{\sqrt{ \ga ( \gx )}} 
\prod'_{\widetilde{C_0}} \left( \frac
{ 1- \mu^{(0)}(\widetilde{C_0} )
 e^{-\sqx l(\widetilde{C_0} )}}
{ 1- \mu^{(0)}(\widetilde{C_0} )
 e^{-\sqg l(\widetilde{C_0} )}} \right)
\prod_{\widetilde{C_0}} \left( \frac
{ 1- \mu(\widetilde{C_0} )
 e^{-\sqg l(\widetilde{C_0} )}}
{ 1- \mu(\widetilde{C_0} )
 e^{-\sqx l(\widetilde{C_0} )}} \right)
\ee
$\fp \prod'_{\widetilde{C_0}} $ means that we only consider primitive orbits
 with reflections in $O$.
 
\vskip1cm

\noindent
As an example, let us see how (\ref{j151})
 works with  the star-graph of Figure 1 with $n=2$ legs. 
 
\noindent
Only one primitive orbit,
 with length $2(l_{01}+l_{02})$ and two reflection
factors equal to $1$, contributes to the product 
 $\fp \prod_{\widetilde{C_0}} $.  

\noindent
Concerning $\fp \prod'_{\widetilde{C_0}} $, two orbits, with lengths $2l_{01}$
 and $2l_{02}$, will contribute. For each one, the two reflection factors
  are $(-1)$ and $1$. For $L$, we get the result:
\bea
L & = & \frac{1}{\sqrt{ \ga ( \gx )}}
 \frac
{ (1 +
 e^{- 2 \sqx l_{01} } )
  (1 +
 e^{- 2 \sqx l_{02} } )}
{ (1 +
 e^{- 2 \sqg l_{01} } )
  (1 +
 e^{- 2 \sqg l_{02} } )} 
 \frac
{  (  1  - 
 e^{- 2 \sqg (  l_{01} +  l_{02} ) } )    }
{  (  1  - 
 e^{- 2 \sqx (  l_{01} +  l_{02} ) } )    } \label{j200} \\
 & = & \frac{1}{\sqrt{ \ga ( \gx )}}
\frac{ \cosh \sqx l_{01} \cosh \sqx l_{02}}
{ \cosh \sqg l_{01} \cosh \sqg l_{02}}
\frac{ \sinh \sqg ( l_{01} + l_{02}   )  }
{ \sinh \sqx ( l_{01} + l_{02}   )  } \label{j201} 
\eea
in agreement with (\ref{j131}).

\vskip.7cm

\noindent
For the ring of Figure 2, two (time-reversed) orbits of length $l$
 should contribute to $\fp \prod_{\widetilde{C_0}} $ and only one of
 length $2l$  should contribute to $\fp \prod'_{\widetilde{C_0}}$. So:
 
\bea
L  & = & \frac{1}{\sqrt{ \ga ( \gx )}}
 \frac
{ (1 - e^{- 2 \sqx l } ) }
{ (1 - e^{- 2 \sqg l } ) }
 \frac
{ (1 - e^{-  \sqg l } )^2 }
{ (1 - e^{-  \sqx l } )^2 }  \label{ring100} \\
  & = & \frac{1}{\sqrt{ \ga ( \gx )}}
  \frac{ \mbox{th}  \sqg   \frac{l}{2} }
       { \mbox{th}  \sqx   \frac{l}{2} }    \label{ring101}
\eea
as given by (\ref{j133b}).

\end{document}